\documentclass[preprint,aps,prc,showpacs,nofootinbib]{revtex4}

\usepackage{epsfig}
\usepackage{graphicx}

\begin{document}

\title{Model independent properties of two-photon exchange in elastic electron proton scattering}

\author{M. P. Rekalo}
\altaffiliation{Permanent address:
\it NSC Kharkov Physical Technical Institute, 61108 Kharkov, Ukraine}
\affiliation{\it DAPNIA/SPhN, CEA/Saclay, 91191 Gif-sur-Yvette Cedex,
France }

\author{E. Tomasi-Gustafsson}
\affiliation{\it DAPNIA/SPhN, CEA/Saclay, 91191 Gif-sur-Yvette Cedex,
France }

\date{\today}

\begin{abstract}
We derive from first principles, as the C-invariance of the electromagnetic interaction and the crossing symmetry, the general properties of two-photon exchange in electron-proton elastic scattering. We show that the presence of this mechanism destroys the linearity of the Rosenbluth separation.
\end{abstract}

\maketitle
\section{Introduction}
Recent developments in the field of hadron electromagnetic form factors (FFs) are due to the very precise and surprising data obtained at the Jefferson Laboratory (JLab), in $\vec e+p\to e+\vec p$ elastic scattering, based on the polarization transfer method \cite{Jo00,Ga02}, which show that the electric and magnetic distributions in the proton are different.

The application of the polarization transfer method, proposed about 30 years ago \cite{Re68} has been possible only recently, as it needs high intensity polarized beams, large solid angle spectrometers and advanced techniques of polarimetry in the GeV range. Experiments have been performed at JLab up to $Q^2=5.6$ GeV$^2$ and an extension up to 9 GeV$^2$ is in preparation \cite{00111}. 

The existing data show a discrepancy between the $Q^2$-dependence of the ratio
$R= \mu_p G_{Ep}$/$G_{Mp}$ of the electric to the magnetic proton form factors ($Q^2$ is the momentum transfer squared, $\mu_p$=2.79 is the proton magnetic moment), whether derived with the standard Rosenbluth separation 
\cite{Ar75} or with the polarization method. 

Therefore a careful experimental and theoretical analysis of this problem is necessary. The important point here is the calculation of radiative corrections to the differential cross section and to polarization observables in elastic $eN$-scattering. If these corrections are large (in absolute value) for the differential cross section \cite{Mo69}, in particular for high resolution experiments, a simplified estimation of radiative corrections to polarization phenomena \cite{Ma00} shows that radiative corrections are small for the ratio $P_L/P_T$ of longitudinal to transverse polarization of the proton emitted in the elastic collision of polarized electrons with an unpolarized proton target.

For this reaction, the one-photon exchange is considered to be the main mechanism. In the standard calculations \cite{Mo69}, the two-photon exchange mechanism is only partially taken into account considering the special part of the integral, where one photon carries all the momentum transfer and the second photon is almost real. This contribution allows to overcome the problem of the 'infrared' divergence. But it has been pointed out \cite{Gu73}  that, at large momentum transfer, the role of another mechanism, where the momentum transfer is shared between the two photons, can be relatively increased, due to the steep decreasing of the electromagnetic form factors with $Q^2$. This effect can eventually become so large that the traditional description of the electron-hadron interaction in terms of electromagnetic currents (and electromagnetic form factors) can become incorrect.

Numerous tests of the validity of the one-photon mechanism have been done in the past, using different methods: test of the linearity of the Rosenbluth formula for the differential cross section, comparison of the $e^+p$ and $e^-p$-cross sections, attempts to measure various T-odd polarization observables.

Note that the two-photon exchange should appear at smaller $Q^2$ for heavier targets: $d$, $^3\!He$,  $^4\!He$, because the corresponding form factors decrease faster with $Q^2$ in comparison with protons. In \cite{Re99} the possible effects of $2\gamma$-exchange have been estimated from  the precise data on the structure function $A(Q^2)$, obtained at Jlab in electron deuteron elastic scattering, up to $Q^2=6$ GeV$^2$ \cite{Al99,Ab99}. The possibility of $2\gamma$-corrections has not been excluded by this analysis, starting from $Q^2=1$ GeV$^2$, and the necessity of  dedicated experiments was pointed out. From this kind of consideration, one would expect to observe the two-photon contribution in $eN$-scattering at larger momentum transfer, for $Q^2\simeq 10$ GeV$^2$.
 
 The exact calculation of the $2\gamma$-contribution to the amplitude of the $e^{\pm} p\to e^{\pm} p$-process requires the knowledge of the matrix element for the double virtual Compton scattering, $\gamma^*+N\to\gamma^*+N$, in a large kinematical region of colliding energy and virtuality of both photons, and can not be done in a model independent form. 
 
However general properties of the hadron electromagnetic interaction, as the C-invariance and the crossing symmetry, give rigorous prescriptions for different observables for the elastic scattering of electrons and positrons by nucleons, in particular for the differential cross section and for the proton polarization, induced by polarized electrons. These concrete prescriptions help in identifying a possible manifestation of the two-photon exchange mechanism. For example, an attempt \cite{Gu03} of resolving the discrepancy between the existing data on the ratio $R$, conserving the linear $\epsilon$-dependence of the elastic cross section in presence of  $2\gamma$-corrections is in contradiction with the C-invariance of the electromagnetic interaction ($\epsilon$ is the degree of polarization for the virtual photon).

The purpose of this paper is to derive the correct $\epsilon$-dependence of the 
$2\gamma$-contribution to the differential cross section and to find a 'model independent' parametrization of these additional terms.  The experimental test of the predicted $\epsilon$-dependence of the differential cross section will be a signature of the presence of the $2\gamma$-contribution and allow to estimate its role.

\section{Crossing symmetry and C-invariance}

The standard expression of the matrix element for elastic $eN$-scattering, in framework of one-photon exchange mechanism, is:

\begin{equation}
{\cal  M}_1 =\displaystyle\frac {e^2}{Q^2}\overline{u}(k_2)\gamma_{\mu}u(k_1) \overline{u}(p_2)\left [F_{1N}(Q^2)\gamma_{\mu}-
\displaystyle\frac{\sigma_{\mu\nu}q_{\nu}}{2m}F_{2N}(Q^2)\right] u(p_1),
\label{eq:mat}
\end{equation}
where $k_1$ $(p_1)$ and $k_2$ $(p_2)$ are the four-momenta of the initial and final electron (nucleon), $m$ is the nucleon mass, $q=k_1-k_2$, $Q^2=-q^2>0$. $F_{1N}$ and $F_{2N}$ are the Dirac and Pauli nucleon electromagnetic form factors, which are real functions of the variable $Q^2$ - in the space-like region of momentum transfer. The same form factors describe also the one-photon mechanism for the elastic scattering of positrons by nucleons. From Eq. (\ref{eq:mat}) one can find the following expression for the differential cross section (in the laboratory  system (Lab)):
\begin{equation}
\displaystyle\frac{d\sigma}{d\Omega}_e=\sigma_M\left [ G_{MN}^2(Q^2)+
\displaystyle\frac{\epsilon}{\tau}G_{EN}^2(Q^2)\right ],
\label{eq:csst}
\end{equation}
$$\tau=\displaystyle\frac{Q^2}{4m^2},~G_{MN}=F_{1N}+F_{2N},~G_{EN}=F_{1N}-\tau F_{2N}$$
where $\sigma_M$ is the Mott cross section, for the scattering of unpolarized electrons by a point charge particle (with spin 1/2), 
$\epsilon$ is another independent kinematical variable, which, together with $Q^2$,  fully determines the kinematics of  elastic $eN$-scattering and can be written, the limit of $m_e=0$, as:
\begin{equation}
\epsilon=
\displaystyle\frac{1}{1+2(1+\tau)\tan^2 \displaystyle\frac{\theta_e}{2}},
\label{eq:csst1}
\end{equation}
where $\theta_e$ is the electron scattering angle in Lab system. Therefore 
$0 (\theta_e=\pi)\le \epsilon\le 1 (\theta_e=0)$. 

If one takes into account the two-photon mechanism, the expressions of the matrix element, Eq. (\ref{eq:mat}), and of the differential cross section, Eq. (\ref{eq:csst}), are essentially modified.

It requires, first of all, a generalization of the spin structure of the matrix element, which can be done, in analogy with elastic $np$-scattering \cite{Go57}, using the general properties of the electron-hadron interaction, such as the P-invariance and the relativistic invariance. Taking into account the identity of the initial and final states and the T-invariance of the electromagnetic interaction, the processes $e^{\pm}N\to e^{\pm} N$, in which  four  particles with spin 1/2 participate, are characterized by six independent products of four-spinors, describing the initial and final fermions.  The corresponding (model independent) parametrization of the matrix element can be done in many different but equivalent forms, in terms of six invariant complex amplitudes, ${\cal A}_i(s,Q^2)$, $i=1-6$, which are functions of two independent variables,  and  $s=(k+p_1)^2$ is the square of the total energy of the colliding particles. In the physical region of the reaction $e^{\pm} N\to e^{\pm} N$ the conditions: $Q^2\ge 0$ and $s\ge(m+m_e)^2\simeq m^2$, apply. 

Previously, another set of variables, $\epsilon$ and $Q^2$, which is equivalent  to $s$ and $Q^2$ (in Lab system) was considered. The variables $\epsilon$ and $Q^2$ are well adapted to the description of the properties of one-photon exchange for elastic $eN$-scattering, because, in this case, only the $Q^2$-dependence of the form factors has a dynamical origin, whereas the linear $\epsilon$-dependence in Eq. (\ref{eq:csst}) is a trivial consequence of the one-photon mechanism. On the other hand, the variables $s$ and $Q^2$ are
better suited to the analysis of the implications from crossing symmetry\footnote{The concept of crossing symmetry was introduced by M. Gell-Mann and M.L. Goldberger \protect\cite{Ga66}, and was successfully applied not only in QED, but in the analysis of different processes, induced by the strong and electromagnetic interaction.}.

The conservation of the lepton helicity, which is a general property of the electromagnetic interaction in electron-hadron scattering, reduces the number of invariant amplitudes for elastic $eN$-scattering, in general complex functions of $s$ and $Q^2$, from six to three.

Therefore we can write the following general parametrization  of the spin structure of the matrix element for elastic $eN$-scattering, following the formalism of \cite{Go57}:
\begin{equation}
{\cal  M}=\displaystyle\frac{e^2}{Q^2}\overline{u}(k_2)\gamma_{\mu}u(k_1)
\overline{u}(p_2) \left [{\cal  A}_1(s,Q^2)\gamma_{\mu}-{\cal  A}_2(s,Q^2) 
\displaystyle\frac{\sigma_{\mu\nu}q_{\nu}}{2m}+{\cal  A}_3(s,Q^2)\hat K {\cal  P}_{\mu}\right] u(p_1),
\label{eq:mat1}
\end{equation}
$$K=\displaystyle\frac{k_1+k_2}{2},~{\cal  P}=\displaystyle\frac{p_1+p_2}{2},$$
where ${\cal  A}_1-{\cal  A}_3$ are the corresponding invariant amplitudes. 

In case of one-photon exchange 
$$
{\cal  A}_1(s,Q^2)\to F_{1N}(Q^2),~
{\cal  A}_2(s,Q^2)\to F_{2N}(Q^2),~
{\cal  A}_3\to 0.
$$
But in the general case (with multi-photon exchanges) the situation is more complicated, because:
\begin{itemize}
\item The amplitudes ${\cal  A}_i(s,Q^2)$, $i=1-3$, are complex functions of two independent variables, $s$, and $Q^2$. 
\item The set of amplitudes ${\cal  A}_i^{(-)}(s,Q^2)$ for the process $e^-+N\to e^-+N$ is different from the set ${\cal  A}_i^{(+)}(s,Q^2)$ of corresponding amplitudes for positron scattering,  $e^++N\to e^++N$, which means that the properties of the scattering of positrons can not be derived from  ${\cal  A}_i^{(-)}(s,Q^2)$, as in case of the one-photon mechanism.
\item The connection of the amplitudes  ${\cal  A}_i(s,Q^2)$ with the nucleon electromagnetic form factors, $F_{iN}(Q^2)$, is non-trivial, because these amplitudes depend on quantities, as, for example, the form factors of the $\Delta$-excitation - through the amplitudes of the virtual Compton scattering.
\end{itemize}

In this framework, the simple and transparent phenomenology of electron-hadron physics does not hold anymore, and in particular, it would  be very difficult  to extract information on the internal structure of a hadron in terms of electromagnetic form factors, which are real functions of one variable, from electron scattering experiments.

In the following text, we will show that the situation is not so involved, and that even in case of two-photon exchange, one can still use the formalism of form factors, if one takes into account the C-invariance of the electromagnetic interaction of hadrons.

A deeper analysis of Eq. (\ref{eq:mat1}) shows that the spin structure of 
${\cal  A}_1$ and ${\cal  A}_2$ corresponds to exchange by vector particle (in $t$-channel), whereas the spin structure for the amplitude ${\cal  A}_3$ corresponds to tensor exchange. In case of $e^{\pm}N $--elastic scattering, in the $1\gamma+2\gamma$ approximation,  one can write the amplitudes 
${\cal  A}^{(\pm)}_{1,2}(s,Q^2)$ in the following fom:
$${\cal  A}^{(\pm)}_{1,2}(s,Q^2)=\pm F_{1,2N}(Q^2) +
\Delta {\cal  A}^{(\pm)}_{1,2}(s,Q^2),$$
$$\Delta {\cal  A}^{(+)}_{1,2}(s,Q^2)=\Delta {\cal  A}^{(-)}_{1,2}(s,Q^2)\equiv 
\Delta {\cal  A}_{1,2}(s,Q^2)$$
$${\cal  A}^{(+)}_3(s,Q^2)={\cal  A}^{(-)}_3(s,Q^2)\equiv {\cal  A}_3(s,Q^2),$$
where the superscript $(\pm)$ corresponds to $e^{(\pm)} $--scattering. The amplitudes $ \Delta {\cal  A}_{1,2}(s,Q^2)$ and ${\cal  A}_3(s,Q^2)$ contain only the $2\gamma$-contribution, and are equal for  $e^{(\pm)}$--scattering; $ \Delta {\cal  A}_{1,2}$ and ${\cal  A}_3$ are of the order of $\alpha$, $\alpha={e^2}/(4\pi)=1/137$. Note that the difference in the spin structure of these amplitudes, Eq. (\ref{eq:mat1}), results in specific symmetry properties with respect to the change $x\to -x$ $\left (x=\sqrt{\displaystyle\frac{1+\epsilon}{1-\epsilon}}\right )$:
$$
\Delta {\cal  A}_{1,2}(s,-x)=- \Delta {\cal  A}_{1,2}(s,x)
$$
\begin{equation}
{\cal  A}_3(s,-x)=+ {\cal  A}_3(s,x)
\label{eq:eq5}
\end{equation}
The $x$--odd behavior of  $\Delta {\cal  A}_{1,2}(s,x)$--contributions, corresponding to $2\gamma$-exchange with $C=+1$, must compensate the C-odd character of the two vector-like spin structures, $\gamma_{\mu}$ and $\sigma_{\mu\nu}Q_{\nu}$.

To prove this, let us consider, in addition to C-invariance, crossing symmetry, which allows to connect the matrix elements for the cross-channels: $e^-+N\to e^-+N$, in $s$--channel, and $e^++e^-\to N+\overline{N}$, in $t$--channel. The transformation from $s$- to $t$-channel can be realized by the following substitution:
$$k_2\to -k_2,~p_1\to -p_1.$$
and for the invariant variables:
$$s=(k_1+p_1)^2\to (k_1-p_1)^2,~Q^2=-(k_1-k_2)^2\to -(k_1+k_2)^2=-t.$$
The crossing symmetry states that the same amplitudes ${\cal  A}_i(s,Q^2)$ describe the two channels, when the variables $s$ and $Q^2$ scan the physical region of the corresponding channels. So, if $t\ge 4m^2$ and $-1\le\cos\theta\le 1$ ($\theta$ is the angle of the proton production with respect to the electron three-momentum, in the center of mass (CMS) for $e^++e^-\to N+\overline{N}$), the amplitudes ${\cal  A}_i(t,\cos\theta)$,  $i=1-3$, describe the process $e^++e^-\to p+\overline{p}$.

The C-invariance of the electromagnetic hadron interaction and the corresponding selection rules can be  applied to the annihilation channel and this allows to find specific properties for one and two photon exchanges. Morevoer, on the basis of the crossing symmetry, it is possible to transform in a transparent way these properties for the different observables in $eN$-elastic scattering.

To illustrate this, let us consider firstly the one-photon mechanism for $e^++e^-\to p+\overline{p}$. The conservation of the total angular momentum ${\cal J}$  allows one value, ${\cal J}=1$ , and the quantum numbers of the photon: ${\cal J}^P=1^-$, $C=1$. The selection rules with respect to the C and P-invariances allow two states for 
$e^+e^-$ (and $p\overline{p}$):
\begin{equation}
S=1,~\ell=0 \mbox{~and~} S=1,~\ell=2\mbox{~with~} {\cal J}^P=1^-,
\label{eq:tran}
\end{equation}
where $S$ is the total spin and $\ell$ is the orbital angular momentum. As a result the $\theta$-dependence of the cross section for $e^++e^-\to p+\overline{p}$, in the one-photon exchange mechanism is:
\begin{equation}
\displaystyle\frac{d\sigma}{d \Omega}(e^++e^-\to p+\overline{p})\simeq a(t)+b(t)\cos^2\theta, 
\label{eq:sig}
\end{equation}
where $a(t)$ and $b(t)$ are definite quadratic contributions of $G_{Ep}(t)$ and 
$G_{Mp}(t)$, $a(t),~b(t)\ge 0$ at $t\ge 4m^2$.

Using the kinematical relations:
\begin{equation}
\cos^2\theta=\displaystyle\frac{1+\epsilon }{1-\epsilon}=
\displaystyle\frac{\cot^2{\theta_e/2}}{1+\tau}+1
\label{eq:cot}
\end{equation}
between the variables in the CMS of $e^++e^-\to p+\overline{p}$ and in the LAB system for $e^-+p\to e^-+p$, it appears clearly that the one-photon mechanism generates a linear $\epsilon$-dependence (or  $\cot^2{\theta_e/2}$) of the Rosenbluth differential cross section for elastic $eN$-scattering in Lab system.

Let us consider now the $\cos\theta$-dependence of the $1\gamma\bigotimes 2\gamma$-interference contribution to the differential cross section of  $e^++e^-\to p+\overline{p}$. The spin and parity of the $2\gamma$-states 
is not fixed, in general, but only a positive value of C-parity, $C(2\gamma)=+1$, is allowed.
An infinite number of  states with different quantum numbers can contribute, and their relative role is determined by the dynamics of the process $\gamma^*+\gamma^*\to  p+\overline{p}$, with both virtual photons.

But the $\cos\theta$-dependence of the contribution to the differential cross section for the $1\gamma\bigotimes 2\gamma$-interference can be predicted on the basis of its C-odd nature:
\begin{equation}
\displaystyle\frac{d\sigma^{(int)}}{d \Omega}(e^++e^-\to p+\overline{p})=\cos\theta[c_0(t)+c_1(t)\cos^2\theta+c_2(t)\cos^4\theta+...],
\label{eq:sig3}
\end{equation}
where $c_i(t)$, $i=0,1..$ are real coefficients, which are functions of $t$,  only. This odd $\cos\theta$-dependence is essentially different from the even $\cos\theta$-dependence of the cross section for the one-photon approximation, Eq. (\ref{eq:sig}).

From C-invariance it follows also that:
$$
{\cal  A}_3(t,-\cos\theta)={\cal  A}_3(t,+\cos\theta), 
$$
\begin{equation}
\Delta A_{1,2}(t,-\cos\theta )=-A_{1,2}(t,+\cos\theta),
\label{eq:a3}
\end{equation}
which is equivalent to the symmetry relations  (\ref{eq:eq5}).

It is therefore incorrect to approximate the interference contribution to the differential cross section
(\ref{eq:sig3}) by a linear function in $\cos^2\theta$, because it is in contradiction with the C-invariance of hadronic electromagnetic interaction.
Such approximation can be done only when all coefficients $c_i(t)$ vanish, i.e. in absence of $1\gamma\bigotimes 2\gamma$-interference!

Using Eq. (\ref{eq:sig3}), the crossing symmetry allows to predict the non-trivial $\epsilon$-dependence of the interference contribution to the differential cross section of $eN$-scattering, in the Lab system, $\Delta\sigma$:
\begin{equation}
\Delta\sigma(e^-p\to e^-p)\simeq x f(x^2,Q^2), 
\label{eq:eq12}
\end{equation}
$$f(x^2,Q^2)=c_0(Q^2)+c_1(Q^2)x^2+c_2(Q^2)x^4+...,~
x=\sqrt{\displaystyle\frac{1+\epsilon }{1-\epsilon }}.$$
Again the C-invariance does not allow to approximate the function $x f(x^2,Q^2)$ by a linear $\epsilon$-dependence:
$$x f(x^2,Q^2)\neq d_0(Q^2)+d_1(Q^2)\epsilon,$$
which would make (\ref{eq:eq12}) compatible with the Rosenbluth formula.

Note that the relation:
\begin{equation}
\Delta\sigma(e^-p\to e^-p)= - \Delta\sigma(e^+p\to e^+p)
\label{eq:eq12a}
\end{equation}
is correct, with the evident correlation of $\Delta\sigma$ with the possible deviation of the differential cross section from a linear $\epsilon$-dependence.

\section{Possible quantum numbers of two-nucleon exchange}

Let us now analyze the $\cos\theta$-dependence of the interference terms for the lowest possible values ${\cal J}^P$ for the $2\gamma$-system, in order to get a hint of the relative values of the coefficients $c_i(t)$ in Eq. (\ref{eq:sig3}). Taking into account the conservation of the leptonic and the nucleonic electromagnetic currents, $q\cdot \ell=q\cdot {\cal J}=0$, the CMS spin structure of the one-photon amplitude for the annihilation process $e^++e^-\to p+\overline{p}$ can be written as:
\begin{equation}
{\cal M}_1=\displaystyle\frac{e^2}{t} \ell\cdot{\cal J}=-\displaystyle\frac{e^2}{t} \vec\ell\cdot\vec{\cal J},
\label{eq:eq12b}
\end{equation}
with
\begin{equation}
 \vec\ell=\sqrt{t}\phi^{\dagger}_2(\vec\sigma-\hat{\vec k}\vec\sigma\cdot\hat{\vec k})\phi_1,
\label{eq:eq14}
\end{equation}
\begin{equation}
\vec{\cal J}=\sqrt{t}\chi^{\dagger}_2\left [ G_M(t)(\vec\sigma-
\hat{\vec p}\vec\sigma\cdot\hat{\vec p})+\displaystyle\frac{1}{\sqrt\tau}G_E(t)\hat{\vec p}\vec\sigma\cdot\hat{\vec p} \right ]\chi_1, 
\label{eq:eq15}
\end{equation}
\begin{equation}
G_M(t)=F_1(t)+F_2(t),~G_E(t)=F_1(t)+\tau F_2(t),~
\tau=\displaystyle\frac{t}{4m^2},
\label{eq:eq16}
\end{equation}
where $\phi_1$ and $\phi_2$ ($\chi_1$ and $\chi_2$) are the two-component spinors of the electron and positron (proton and antiproton), 
$\hat{\vec k} $ ($\hat{\vec p} $) is the unit vector along the three momentum of the electron (proton) in CMS.

Note that the term $G_{Mp}(t)-\displaystyle\frac{1}{\sqrt\tau}G_{Ep}(t)$, describes the $\overline{p}p-$production with $\ell$=2. Therefore, at threshold, $\tau\to 1$, where the finite radius of the strong interaction allows the $\overline{p}p$-production only in S-state, the following relation: $G_{EN}(t)=G_{MN}(t),~t\to 4 m^2$ holds and it is the physical background of this so particular relation between the nucleon electromagnetic form factors at threshold.

Summing over the polarizations of the $p\overline{p}$-system and averaging over the polarizations of the initial $e^+e^-$-system, one can find with the help of Eqs. (\ref{eq:eq14},\ref{eq:eq15}):
\begin{equation} 
\overline{|\vec\ell\cdot\vec{\cal J}|^2}=\displaystyle\frac{t}{2}\left [(1+\cos^2\theta)|G_{Mp}(t)|^2+
\displaystyle\frac{1}{\tau}\sin^2\theta |G_{Ep}(t)|^2\right]
\label{eq:eq17}
\end{equation}
with the standard $\theta$-dependence of the differential cross section for $e^++e^-\to\overline{p} +p$ \cite{Zi62}, calculated in frame of the one-photon mechanism.

After substituting $t\to Q^2$ and $\cos\theta^2\to (1+\epsilon)/(1-\epsilon)$ in 
Eq. (\ref{eq:eq17}), one can find the linear $\epsilon$-dependence for the Rosenbluth formula for the differential cross section of elastic $eN$-scattering in terms of $|G_{Ep}|^2$ and $|G_{Mp}|^2$ in Lab system.

In the same way one can find the two-component spin structure for the $A_3$-contribution to the matrix element for $e^++e^-\to\overline{p}+p$, using Eq. \ref{eq:mat1}:
\begin{equation}
\overline{u}(-k_2)\hat {\cal P} u(k_1)\overline{u}(p_2) \hat{\cal K} u(-p_1)={\cal L}{\cal N},
\label{eq:eq18}
\end{equation}
\begin{equation}
{\cal L}=\displaystyle\frac{m}{2}\sqrt{t(\tau-1)}\phi^{\dagger}_2(\vec\sigma\cdot\hat{\vec p}-\cos\theta\vec\sigma\cdot\hat{\vec k})\phi_1,
\label{eq:eq19}
\end{equation}
\begin{equation}
{\cal N}=-\displaystyle\frac{t}{2}\chi^{\dagger}_2(\vec\sigma\cdot\hat{\vec k}-\cos\theta\vec\sigma\cdot\hat{\vec p}+ \displaystyle\frac{1}{\sqrt\tau}\cos\theta\vec\sigma\cdot\hat{\vec p})\chi_1.
\label{eq:eq20}
\end{equation}
The corresponding interference term can be written as:
\begin{equation}
\overline{\vec\ell\cdot\vec{\cal J}{\cal L}^*{\cal N}^*}\simeq Re\left [G_M(t)-\displaystyle\frac{1}{\tau}G_E(t)\right ]\cos\theta\sin^2\theta
\label{eq:eq21}
\end{equation}
with a specific $\theta$-dependence. Applying the crossing symmetry, this $\theta$-dependence (from the interference term) generates a definite $\epsilon$-dependence of the corresponding contribution to the differential cross section of the $eN$-scattering in Lab system:
\begin{equation}
1\gamma\bigotimes 2\gamma\simeq \displaystyle\frac{2\epsilon}{1-\epsilon}
\sqrt{\displaystyle\frac{1+\epsilon}{1-\epsilon}},
\label{eq:eq21a}
\end{equation}
which is essentially non linear, in contrast with the assumptions done in \cite{Gu03,Bl03}. If the interference term had a linear $\epsilon$-dependence,  the product $\sqrt{\displaystyle\frac{1+\epsilon}{1-\epsilon}}f(x^2),$ with $x^2=\displaystyle\frac{1+\epsilon}{1-\epsilon}$, would be  $\epsilon$-independent, again in contradiction with the C-invariance of hadronic electromagnetic interaction and with crossing symmetry.

Let us discuss now how unique is the $\cos\theta\sin^2\theta$-dependence for $e^++e^-\to p+\overline{p}$. One can show, on the basis of Eqs. (\ref{eq:eq19}) and (\ref{eq:eq20}), that such term arises from a definite superposition of states of the $2\gamma$-system with quantum numbers 
${\cal J}^P=1^+$ and 2$^+$, when the $e^+e^-$-system has $S=\ell=1$. The individual states have different structures:
\begin{eqnarray*}
&{\cal J}^P=1^+, \ell=1 &\to \cos\theta Re G_M(t),\\
&{\cal J}^P=2^+, \ell=1 &\to \cos\theta [Re G_M(t)+\sin^2\theta\displaystyle\frac{1}{\tau} Re G_E(t)].
\end{eqnarray*}
The simplest linear $\cos\theta$-dependence corresponds to the exchange by the axial state with 
${\cal J}^P=1^+,$ $\ell=S=1$. It is therefore possible to use, for the discussion of interference phenomena instead of (\ref{eq:mat1}), another equivalent parametrization of $2\gamma$-exchange:
$${\cal M}_2\simeq \tilde{\cal A}_3(s,Q^2)\overline{u}(k_2)\gamma_{\mu}\gamma_5u(k_1)
\overline{u}(p_2)\gamma_{\mu}\gamma_5u(p_1).$$
%%%%%%%%%%%%%%%%%%%%%%%%%
\section{Conclusions}
%%%%%%%%%%%%%%%%%%%%%%%%%

The general symmetry properties of electromagnetic interaction, such as the C-invariance and the crossing symmetry, allow to obtain  rigorous results concerning two-photon exchange contributions for elastic $eN$-scattering and to analyze the effects of this mechanism in $eN$-phenomenology.

The form factors $G_{EN}(Q^2)$ and $G_{MN}(Q^2)$ and the $2\gamma$-amplitudes, 
${\cal A}_3(s,Q^2)$ and  $\Delta {\cal A}_{1,2}(s,Q^2)$are the same for $e^+p$ and $e^-p$ elastic scattering. This allows to prove that the sum of the differential cross sections for $e^{\pm} p$-interaction has the standard  Rosenbluth dependence on the nucleon form factors.

The $\epsilon$-dependence of the interference contribution to the differential cross section of $e^{\pm}p$ elastic scattering is very particular. Any approximation of this term by a linear function in the variable $\epsilon$ is in contradiction with C-invariance and crossing symmetry of the electromagnetic interaction. 

The formal expression of the $\epsilon$-dependence of the interference contribution depends on the quantum numbers of the $2\gamma$-system.

To have a quantitative estimation of the relative role of two-photon physics in $eN$-interaction, it is necessary to measure the $\epsilon$-dependence of the differential cross section of $eN$ elastic scattering in several points,  and study this behavior in terms of the specific variable $\sqrt{(1+\epsilon)/(1-\epsilon)}$. This will be the unambiguous signature of two-photon contributions. 
A similar analysis can be done for polarization phenomena, and it is the object of a future paper.

We thank N. Kochelev, J. Arrington and L. Pentchev for a careful reading of the manuscript and for pointing out misprints along the paper.
{}

\end{document}